\documentclass[12pt]{article}
\usepackage{a4wide}
\begin{document}

\mbox{}\hfill AEI--2005--134\\
\mbox{}\hfill IGPG--06/3--1\\
\mbox{}\hfill gr-qc/0603110

{\renewcommand{\thefootnote}{\fnsymbol{footnote}}
\begin{center}
{\LARGE  Quantum Cosmology\footnote{Encyclopedia of Mathematical Physics, 
eds.\ J.-P.~Fran\c coise, G.~L.~Naber and
Tsou S.~T.,  Oxford: Elsevier, 2006 (ISBN 978-0-1251-2666-3), 
volume 4, page 153.}}\\
\vspace{1.5em}
Martin Bojowald\footnote{e-mail address: {\tt bojowald@gravity.psu.edu}}
\\
\vspace{0.5em}
Institute for Gravitational Physics and Geometry, The Pennsylvania
State University, University Park, PA 16802, USA\\[0.5em]
and\\[0.5em]
Max-Planck-Institut f\"ur Gravitationsphysik, Albert-Einstein-Institut,\\
Am M\"uhlenberg 1, D-14476 Potsdam, Germany\\

\vspace{1.5em}
\end{center}
}

\setcounter{footnote}{0}

{\bf Synopsis:}
Quantum cosmology in general denotes the application of quantum physics
to the whole universe and thus gives rise to many realizations and
examples, covering problems at different mathematical and conceptual
levels. It is related to quantum gravity and more specifically
describes the application to cosmological situations rather than the
construction and analysis of quantum field equations. As there are
several different approaches to quantum gravity, equations for quantum
cosmology are not unique. Most investigations have been performed in
the context of canonical quantization, where Wheeler--DeWitt like
equations are the prime object. Applications are mostly conceptual,
ranging from possible resolutions of classical singularities and
explanations of the uniqueness of the universe to the origin of seeds
for a classical world and its initial conditions.


\section{Introduction}

Classical gravity, through its attractive nature, leads to high
curvature in important situations. In particular, this is realized in
the very early universe where in the backward evolution energy
densities are growing until the theory breaks down. Mathematically,
this point appears as a singularity where curvature and physical
quantities diverge and the evolution breaks down. It is not possible
to set up an initial value formulation at this place in order to
determine the further evolution.

In such a regime, quantum effects are expected to play an important
role and to modify the classical behavior such as the attractive
nature of gravity or the underlying space-time structure. Any
candidate for quantum gravity thus allows to re-analyze the
singularity problem in a new light which implies tests of
characteristic properties of the respective candidate. Moreover, close
to the classical singularity in the very early universe quantum
modifications will give rise to new equations of motion which turn
into Einstein's equations only on larger scales. The analysis of these
equations of motion leads to new classes of early universe
phenomenology.

The application of quantum theory to cosmology presents a unique
problem with not only mathematical but also many conceptual and
philosophical ramifications. Since by definition there is only one
universe which contains everything accessible, there is no place for
an outside observer separate from the quantum system. This eliminates
the most straightforward interpretations of quantum mechanics and
requires more elaborate, and sometimes also more realistic,
constructions such as decoherence. From the mathematical point of
view, this situation is often expected to be mirrored by a new type of
theory which does not allow one to choose initial or boundary
conditions separately from the dynamical laws. Initial or boundary
conditions, after all, are meant to specify the physical system
prepared for observations which is impossible in cosmology. Since we
observe only one universe, the expectation goes, our theories should
finally present us with only one, unique solution without any freedom
for further conditions. This solution then contains all information
about observations as well as observers. Mathematically, this is an
extremely complicated problem which has received only scant
attention. Equations of motion for quantum cosmology are usually of
the type of partial differential or difference equations, such that
new ingredients from quantum gravity are needed to restrict the large
freedom of solutions.

\section{Minisuperspace approximation}

In most investigations, the problem of applying full quantum gravity
to cosmology is simplified by a symmetry reduction to homogeneous or
isotropic geometries. Originally, the reduction was performed at the
classical level, leaving in the isotropic case only one gravitational
degree of freedom given by the scale factor $a$. Together with
homogeneous matter fields, such as a scalar $\phi$, there are then
only finitely many degrees of freedom which one can quantize using
quantum mechanics. The classical Friedmann equation for the evolution
of the scale factor, depending on the spatial curvature $k=0$ or
$\pm1$, is then quantized to the Wheeler--DeWitt equation, commonly
written as
\begin{equation} \label{WdW}
 \left(\frac{1}{9}\ell_{\rm P}^4a^{-x} \frac{\partial}{\partial a} a^x
 \frac{\partial}{\partial a}- ka^2\right) \psi(a,\phi)=-\frac{8\pi
 G}{3} a\hat{H}_{\rm matter}(a) \psi(a,\phi)
% \left[a^{-x}\frac{\partial}{\partial a}a^x\frac{\partial}{\partial a}
%-\frac{3}{4\pi G}a^{-2} \frac{\partial^2}{\partial\phi^2}-
%\frac{9\pi^2}{4G^2}(ka^2-\frac{8\pi
%G}{3}a^4V(\phi))\right]\psi(a,\phi)=0
\end{equation}
for the wave function $\psi(a,\phi)$. The matter Hamiltonian
$\hat{H}_{\rm matter}(a)$, such as
\begin{equation}
 \hat{H}_{\rm matter}(a)=-\frac{1}{2}\hbar^2a^{-3}
 \frac{\partial^2}{\partial\phi^2}+a^3V(\phi)\,,
\end{equation}
is left unspecified here, and $x$ parameterizes factor ordering
ambiguities (but not completely). The Planck length $\ell_{\rm
P}=\sqrt{8\pi G\hbar}$ is defined in terms of the gravitational
constant $G$ and the Planck constant $\hbar$.

The central conceptual issue then is the generality of effects seen in
such a symmetric model and its relation to the full theory of quantum
gravity. This is completely open in the Wheeler--DeWitt form since the
full theory itself is not even known. On the other hand, such
relations are necessary to value any potential physical statement
about the origin and early history of the universe. In this context,
symmetric situations thus present models, and the degree to which they
approximate full quantum gravity remains mostly unknown. There are
examples, for instance of isotropic models in anisotropic but still
homogeneous models, where a minisuperspace quantization does not agree
at all with information obtained from the less symmetric model. But
often those effects already have a classical analog such as
instability of the more symmetric solutions. A wider investigation of
the reliability of models and when correction terms from ignored
degrees of freedom have to be included has not been done yet.

With candidates for quantum gravity being available, the current
situation has changed to some degree. It is then not only possible to
reduce classically and then simply use quantum mechanics, but also to
perform at least some of the reduction steps at the quantum level. The
relation to models is then much clearer, and consistency conditions
which arise in the full theory can be made certain to be
observed. Moreover, relations between models and the full theory can
be studied to elucidate the degree of approximation. Even though new
techniques are now available, a detailed investigation of the degree
of approximation given by a minisuperspace model has not been
completed due to its complexity.

This program has mostly been developed in the context of loop quantum
gravity, where the specialization to homogeneous models is known as
loop quantum cosmology. More specifically, symmetries can be
introduced at the level of states and basic operators, where symmetric
states of a model are distributions in the full theory, and basic
operators are obtained by the dual action on those distributions. In
such a way, the basic representation of models is not assumed but
derived from the full theory where it is subject to much stronger
consistency conditions. This has implications even in homogeneous
models with finitely many degrees of freedom, despite the fact that
quantum mechanics is usually based on a unique representation if the
Weyl operators $e^{isq}$ and $e^{itp}$ for the variables $q$ and $p$
are represented weakly continuously in the real parameters $s$ and
$t$.

The continuity condition, however, is not necessary in general,
and so inequivalent representations are possible. In quantum cosmology
this is indeed realized, where the Wheeler--DeWitt representation
assumes that the conjugate to the scale factor, corresponding to
extrinsic curvature of an isotropic slice, is represented through a
continuous Weyl operator, while the representation derived for loop
quantum cosmology shows that the resulting operator is not weakly
continuous. Furthermore, the scale factor has a continuous spectrum in
the Wheeler--DeWitt representation but a discrete spectrum in the loop
representation. Thus, the underlying geometry of space is very
different, and also evolution takes a new form, now given by a
difference equation of the type
\begin{eqnarray} \label{DiffEq}
&&    (V_{\mu+5}-V_{\mu+3})e^{ik}\psi_{\mu+4}(\phi)- (2+k^2)
(V_{\mu+1}-V_{\mu-1})\psi_{\mu}(\phi)\\\nonumber
&&+    (V_{\mu-3}-V_{\mu-5})e^{-ik}\psi_{\mu-4}(\phi)
  = -\frac{4}{3}\pi
G\ell_{\rm P}^2\hat{H}_{\rm matter}(\mu)\psi_{\mu}(\phi)
\end{eqnarray}
in terms of volume eigenvalues $V_{\mu}=(\ell_{\rm
  P}^2|\mu|/6)^{3/2}$. For large $\mu$ and smooth wave functions one
can see that the difference equation reduces to the Wheeler--DeWitt
equation with $|\mu|\propto a^2$ to leading order in derivatives of
$\psi$. At small $\mu$, close to the classical singularity, however,
both equations have very different properties and lead to different
conclusions. Moreover, the prominent role of difference equations
leads to new mathematical problems.

This difference equation is not simply obtained through a
discretization of (\ref{WdW}), but derived from a constraint operator
constructed with methods from full loop quantum gravity. It is thus to
be regarded as more fundamental, with (\ref{WdW}) emerging in a
continuum limit. The structure of (\ref{DiffEq}) depends on properties
of the full theory such that its qualitative analysis allows
conclusions for full quantum gravity.

\section{Applications}

Traditionally, quantum cosmology has focused on three main conceptual
issues:
\begin{itemize}
 \item the fate of classical singularities,
 \item initial conditions and the ``prediction'' of inflation (or
other early universe scenarios),
 \item arrow of time and the emergence of a classical world.
\end{itemize}
The first issue consists of several subproblems since there are
different aspects to a classical singularity. Often, curvature or
energy densities diverge and one can expect quantum gravity to provide
a natural cutoff. More importantly, however, is the fact that the
classical evolution breaks down at a singularity, and quantum gravity,
if it is to cure the singularity problem, has to provide a
well-defined evolution which does not stop. Initial conditions
are often seen in relation to the singularity problem since early
attempts tried to replace the singularity by choosing appropriate
conditions for the wave function at $a=0$. Different proposals then
lead to different solutions for the wave function, whose dependence on
the scalar $\phi$ can be used to determine its probability
distribution such as that for an inflaton. Since initial conditions
often provide special properties early on, the combination of
evolution and initial conditions has been used to find a possible
origin of an arrow of time.

\subsection{Singularities}

While classical gravity is based on space-time geometry and thus
metric tensors, this structure is viewed as emergent only at large
scales in canonical quantum gravity. A gravitational system, such as a
whole universe, is instead described by a wave function which at best
yields expectation values for a metric. The singularity problem thus
takes a different form since it is not metrics which need to be continued as
solutions to Einstein's field equations but the wave function
describing the quantum system. In the strong curvature regime around a
classical singularity one does not expect classical geometry to be
applicable, such that classical singularities may just be a reflection
of the breakdown of this picture, rather than a breakdown of physical
evolution. Nevertheless, the basic feature of a singularity as
presenting a boundary to the evolution of a system equally applies to
the quantum equations. One can thus analyze this issue, using new
properties provided by the quantum evolution.

The singularity issue is not resolved in the Wheeler--DeWitt
formulation since energy densities, with $a$ being a multiplication
operator, diverge and the evolution does not continue anywhere beyond
the classical singularity at $a=0$. In some cases one can formally
extend the evolution to negative $a$, but this possibility is not
generic and leaves open what negative $a$ mean geometrically. This is
different in the loop quantization: here, the theory is based on triad
rather than metric variables. There is thus a new sign factor
corresponding to spatial orientation, which implies the possibility of
negative $\mu$ in the difference equation. The equation is then
defined on the full real line with the classical singularity $\mu=0$
in the interior. Outside $\mu=0$ we have positive volume at both
sides, and opposite orientations. Using the difference equation one
can then see that the evolution does not break down at $\mu=0$,
showing that the quantum evolution is singularity free.

For the example (\ref{DiffEq}) shown here, one can follow the
evolution for instance backward in internal time $\mu$, starting from
initial values for $\psi$ at large positive $\mu$. By successively
solving for $\psi_{\mu-4}$, the wave function at lower $\mu$ is
determined. This goes on in this manner only until the coefficient
$V_{\mu-3}-V_{\mu-5}$ of $\psi_{\mu-4}$ vanishes, which is the case if
and only if $\mu=4$. The value $\psi_0$ of the wave function right at
the classical singularity is thus not determined by initial data, but
one can easily see that it completely drops out of the evolution. In
fact, the wave function at all negative $\mu$ is uniquely determined
by initial values at positive $\mu$. The equation (\ref{DiffEq})
corresponds to one particular ordering, which in the Wheeler--DeWitt
case is usually parameterized by the parameter $x$ (although the
particular ordering obtained from the continuum limit of
(\ref{DiffEq}) is not contained in the special family
(\ref{WdW})). Other non-singular orderings exist, such as that after
symmetrizing the constraint operator in which case coefficients never
become zero.

In more complicated systems, this behavior is highly non-trivial but
still known to be realized in a similar manner. It is not automatic
that the internal time evolution does not continue since even in
isotropic models one can easily write difference equations for which
the evolution breaks down. That the most natural orderings imply
non-singular evolution can be taken as support of the general
framework of loop quantum gravity. It should also be noted that the
mechanism described here, providing essentially a new region beyond a
classical singularity, presents one mechanism for quantum gravity to
remove classical singularities, and so far the only known
one. Nevertheless, there is no claim that the ingredients have to be
realized in any non-singular scenario in the same manner. Different
scenarios can be imagined, depending on how quantum evolution is
understood and what the interpretation of non-singular behavior is. It
is also not claimed that the new region is semiclassical in any sense
when one looks at it at large volume. If the initial values for the
wave function describe a semiclassical wave packet, its evolution
beyond the classical singularity can be deformed and develop many
peaks. What this means for the re-emergence of a semiclassical
space-time has to be investigated in particular models, and also in
the context of decoherence.

\subsection{Initial conditions}

Traditional initial conditions in quantum cosmology have been
introduced by physical intuition. The main mathematical problem, once
such a condition is specified in sufficient detail, then is to study
well-posedness, for instance for the Wheeler--DeWitt equation. Even
formulating initial conditions generally, and not just for isotropic
models, is complicated, and systematic investigations of the
well-posedness have rarely been undertaken. An exception is the
historically first such condition, due to DeWitt, that the wave
function vanish at parts of minisuperspace, such as $a=0$ in the
isotropic case, corresponding to classical singularities. This
condition, unfortunately, can easily be seen to be ill-posed in
anisotropic models where in general the only solution vanishes
identically. In other models, the limit $\lim_{a\to0}\psi(a)$ does not
even exist. Similar problems of the generality of conditions arise in
other scenarios. Most well-known are the no-boundary and tunneling
proposal where initial conditions are still imposed at $a=0$, but with
a non-vanishing wave function there.

This issue is quite different for difference equations since at first
the setup is less restrictive: there are no continuity or
differentiability conditions for a solution. Moreover, oscillations
which become arbitrarily rapid, which can be responsible for the
non-existence of $\lim_{a\to0}\psi(a)$, cannot be supported on a
discrete lattice. It can then easily happen that a difference equation
is well-posed, while its continuum limit with an analogous initial
condition is ill-posed. One example are the dynamical initial
conditions of loop quantum cosmology which arise from the dynamical
law in the following way: The coefficients in (\ref{DiffEq}) are not
always non-zero but vanish if and only if they are multiplied with the
value of the wave function at the classical singularity $\mu=0$. This
value thus decouples and plays no role in the evolution. The instance
of the difference equation that would determine $\psi_0$, e.g.\ the
equation for $\mu=4$ in the backward evolution, instead implies a
condition on the previous two values, $\psi_4$ and $\psi_8$ in the
example. Since they have already been determined in previous iteration
steps, this translates to a linear condition on the initial values
chosen. We thus have one example where indeed initial conditions and
the evolution follow from only one dynamical law, which also extends
to anisotropic models. Without further conditions, the initial value
problem is always well-posed, but may not be complete, in the sense
that it results in a unique solution up to norm. Most of the
solutions, however, will be rapidly oscillating. In order to guarantee
the existence of a continuum approximation, one has to add a condition
that these oscillations are suppressed in large volume regimes. Such a
condition can be very restrictive, such that the issue of
well-posedness appears in a new guise: non-zero solutions do exist,
but in some cases all of them may be too strongly oscillating.

In simple cases one can advantageously use generating function
techniques to study oscillating solutions, at least if oscillations
are of alternating nature between two subsequent levels of the
difference equation. The idea is that a generating function
$G(x)=\sum_n\psi_n x^n$ has a stronger pole at $x=-1$ if $\psi_n$ is
alternating compared to a solution of constant sign. Choosing initial
conditions which reduce the pole order thus implies solutions with
suppressed oscillations. As an example, we can look at the difference
equation
\begin{equation}
 \psi_{n+1}+\frac{2}{n}\psi_n-\psi_{n-1}=0
\end{equation}
whose generating function is
\begin{equation}
 G(x)=\frac{\psi_1x+\psi_0(1+2x(1-\log(1-x)))}{(1+x)^2}\,.
\end{equation}
The pole at $x=-1$ is removed for initial values
$\psi_1=\psi_0(2\log2-1)$ which corresponds to non-oscillating
solutions. In this way, analytical expressions can be used instead of
numerical attempts which would be sensitive to rounding errors.
Similarly, the issue of finding bounded solutions can be studied by
continued fraction methods. This illustrates how an underlying
discrete structure leads to new questions and the application of new
techniques compared to the analysis of partial differential equations
which appear more commonly.

\section{More general models}

Most of the time, homogeneous models have been studied in quantum
cosmology since even formulating the Wheeler--DeWitt equation in
inhomogeneous cases, so-called midisuperspace models, is
complicated. Of particular interest among homogeneous models is the
Bianchi IX model since it has a complicated classical dynamics of
chaotic behavior. Moreover, through the BKL picture the Bianchi IX
mixmaster behavior is expected to play an important role even for
general inhomogeneous singularities. The classical chaos then
indicates a very complicated approach to classical singularities, with
structure on arbitrarily small scales.

On the other hand, the classical chaos relies on a curvature potential
with infinitely high walls, which can be mapped to a chaotic billiard
motion. The walls arise from the classical divergence of curvature,
and so quantum effects have been expected to change the picture, and
shown to do so in several cases.

Inhomogeneous models have mostly been studied in cases, such as
polarized Gowdy models, where one can reformulate the problem as that
of a massless free scalar on flat Minkowski space. The scalar can then
be quantized with familiar techniques in a Fock space representation,
and is related to metric components of the original model in rather
complicated ways. Quantization can thus be performed, but
transforming back to the metric at the operator level and drawing
conclusions is quite involved. The main issue of interest in the
recent literature has been the investigation of field theory aspects
of quantum gravity in a tractable model. In particular, it turns out
that self-adjoint Hamiltonians, and thus unitary evolution, do not
exist in general.
%  {\bf see gr-qc/0206083, Refs.\ [1]--[11]}

Loop quantizations of inhomogeneous models are available even in cases
where a reformulation as a field theory on flat space does not exist,
or is not being made use of to avoid special gauges. This is
quite valuable in order to see if specific features exploited in
reformulations lead to artifacts in the results. So far, the dynamics
has not been investigated in detail, even though conclusions for the
singularity issue can already be drawn.

From a physical perspective, it is most important to introduce
inhomogeneities at a perturbative level in order to study implications
for cosmological structure formation. On a homogeneous background one
can perform a mode decomposition of metric and matter fields and
quantize the homogeneous modes as well as amplitudes of higher
modes. Alternatively, one can first quantize the inhomogeneous system
and then introduce the mode decomposition at the quantum level.  This
gives rise to a system of infinitely many coupled equations of
infinitely many variables, which needs to be truncated e.g.\ for
numerical investigations. At this level, one can then study the
question to which degree a given minisuperspace model presents a good
approximation to the full theory, and where additional correction
terms should be introduced. It also allows to develop concrete models
of decoherence, which requires a ``bath'' of many weakly interacting
degrees of freedom usually thought of as being provided by
inhomogeneities in cosmology, and an understanding of the
semiclassical limit.

\section{Interpretations}

Due to the complexity of full gravity, investigations without symmetry
assumptions or perturbative approximations usually focus on conceptual
issues. As already discussed, cosmology presents a unique situation
for physics since there cannot be any outside observer. While this
fact has already implications on the interpretation of observations
at the classical level, its full force is noticed only in quantum
cosmology. Since some traditional interpretations of quantum mechanics
require the role of observers outside the quantum system, they do not
apply to quantum cosmology.

Sometimes, alternative interpretations such as Bohm theory or many
world scenarios are championed in this situation, but more
conventional relational pictures are most widely adopted. In such an
interpretation, the wave function yields relational probabilities
between degrees of freedom rather than absolute probabilities for
measurements done by one outside observer. This has been used, for
instance, to determine the probability of the right initial conditions
for inflation, but it is marred by unresolved interpretational issues
and still disputed. Those problems can be avoided by using effective
equations, in analogy to an effective action, which modify classical
equations on small scales. Since the new equations are still of
classical type, i.e.\ differential equations in coordinate time, no
interpretational issues arise at least if one stays in semiclassical
regimes. In this manner, new inflationary scenarios motivated from
quantum cosmology have been developed.

In general, a relational interpretation, though preferable
conceptually, leads to technical complications since the situation is
much more involved and evolution is not easy to disentangle. In
cosmology, one often tries to single out one degree of freedom as
internal time with respect to which evolution of other degrees of
freedom is measured. In homogeneous models one can simply take the
volume as internal time, such as $a$ or $\mu$ earlier, but in full no
candidate is known. Even in homogeneous models, the volume is not
suitable as internal time to describe a possible recollapse. One can
use extrinsic curvature around such a point, but then one has to
understand what changing the internal time in quantum cosmology
implies, i.e.\ whether evolution pictures obtained in different
internal time formulations are equivalent to each other.

There are thus many open issues at different levels, which strictly
speaking do not apply only to quantum cosmology but to all of
physics. After all, every physical system is part of the universe, and
thus a potential ingredient of quantum cosmology. Obviously, physics
works well in most situations without taking into account its being
part of one universe. Similarly, much can be learned about a quantum
universe if only some degrees of freedom of gravity are considered as
in mini- or midisuperspace models. Also complicated interpretational
issues, as important as they are for a deep understanding of quantum
physics, do not prevent the development of physical applications in
quantum cosmology, just as they did not do so in the early stages of
quantum mechanics.

\end{document}